\documentclass[twocolumn]{aastex62}

\newcommand{{\maxi}}{{\it MAXI}}
\newcommand{{\swift}}{{\it Swift}}
\newcommand{\srcname}{{MAXI J1820$+$070 }}

\graphicspath{{./}{figures/}}

\shorttitle{Monitoring State Transitions of MAXI J1820$+$070}
\shortauthors{Shidatsu et al.}

\begin{document}

\title{X-ray and Optical Monitoring of State Transitions in MAXI J1820$+$070}

\correspondingauthor{Megumi Shidatsu}
\email{shidatsu.megumi.wr@ehime-u.ac.jp}

\author[0000-0001-8195-6546]{Megumi Shidatsu}
\affil{Department of Physics, Ehime University, 
2-5, Bunkyocho, Matsuyama, Ehime 790-8577, Japan}
\author{Satoshi Nakahira}
\affil{High Energy Astrophysics Laboratory, RIKEN, 2-1, Hirosawa, Wako, Saitama 351-0198, Japan}
\author{Katsuhiro L. Murata}
\affil{Department of Physics, Tokyo Institute of Technology, 2-12-1 Ookayama, Meguro-ku, Tokyo 152-8551, Japan}
\author{Ryo Adachi}
\affil{Department of Physics, Tokyo Institute of Technology, 2-12-1 Ookayama, Meguro-ku, Tokyo 152-8551, Japan}
\author{Nobuyuki Kawai}
\affil{Department of Physics, Tokyo Institute of Technology, 2-12-1 Ookayama, Meguro-ku, Tokyo 152-8551, Japan}
\author[0000-0001-7821-6715]{Yoshihiro Ueda}
\affil{Department of Astronomy, Kyoto University, Kitashirakawa-Oiwake-cho, Sakyo-ku, Kyoto, Kyoto 606-8502, Japan}
\author{Hitoshi Negoro}
\affil{Department of Physics, Nihon University, 1-8-14 Kanda-Surugadai, Chiyoda-ku, Tokyo 101-8308, Japan}


\begin{abstract}
We report results from the X-ray and optical monitoring of 
the black hole candidate MAXI J1820$+$070 ($=$ASSASN-18ey)
over the entire period of its outburst from March to October 2018.
In this outburst, the source exhibited two sets of ``fast rise and 
slow decay''-type long-term flux variations. 
We found that the 1--100 keV luminosities at two peaks 
were almost the same, although a significant spectral 
softening was only seen in the second flux rise.
This confirms that the state transition from the low/hard 
state to the high/soft state is not determined by the mass 
accretion rate alone. The X-ray spectrum was reproduced 
with the disk blackbody emission and its Comptonization, 
and the long-term spectral variations seen in this 
outburst were consistent with a disk truncation model. 
The Comptonization component, with a photon index 
of 1.5--1.9 and electron temperature of $\gtrsim 40$ keV, 
was dominant during the low/hard state periods, and  
its contribution rapidly decreased (increased) 
during the spectral softening (hardening). 
During the high/soft state period, in which the X-ray spectrum 
became dominated by the disk blackbody component,
the inner disk radius was almost constant, suggesting 
that the standard disk was present down to the inner most 
stable circular orbit. The long-term evolution 
of optical and X-ray luminosities and their correlation 
suggest that the jets substantially contributed to the 
optical emission in the low/hard state, while they 
are quenched and the outer disk emission dominated 
the optical flux in the intermediate state and the high/soft state.

\end{abstract}


\section{Introduction} \label{sec:intro}

The Galactic black hole candidate MAXI J1820$+$020/\\ASASSN-18ey, 
discovered in optical by the All-Sky Automated Survey for 
SuperNovae (ASSAS-SN) project \citep{sha14} 
on 2018 March 6 \citep{tuc18} and in X-rays with 
{\maxi} \citep{mat09} on 2018 March 11 \citep{kaw18a}, 
has been an ideal target to probe black hole accretion 
flows and jets over a wide range of mass accretion rates. 
Thanks to its relatively small distance ($3 \pm 1$ kpc, \citealt{gan18b}) 
and low Galactic extinction ($N_\mathrm{H} = 1.5 \times 10^{21}$ cm$^{-2}$, 
\citealt{utt18}), the source 
has been intensively observed at various wavelengths, with 
different methods (photometry, spectroscopy, and polarimetry), 
and on various time resolutions, from ms to $\gtrsim$day 
\citep[][and many other reports in the Astronomer's 
Telegram]{ken18a,ken18b,ken18c,bag18a,bri18a,lit18,vel18,
utt18,bah18,gar18,sak18,del18,pai18,gan18a,tru18}.

\begin{figure*}[th!]
\plotone{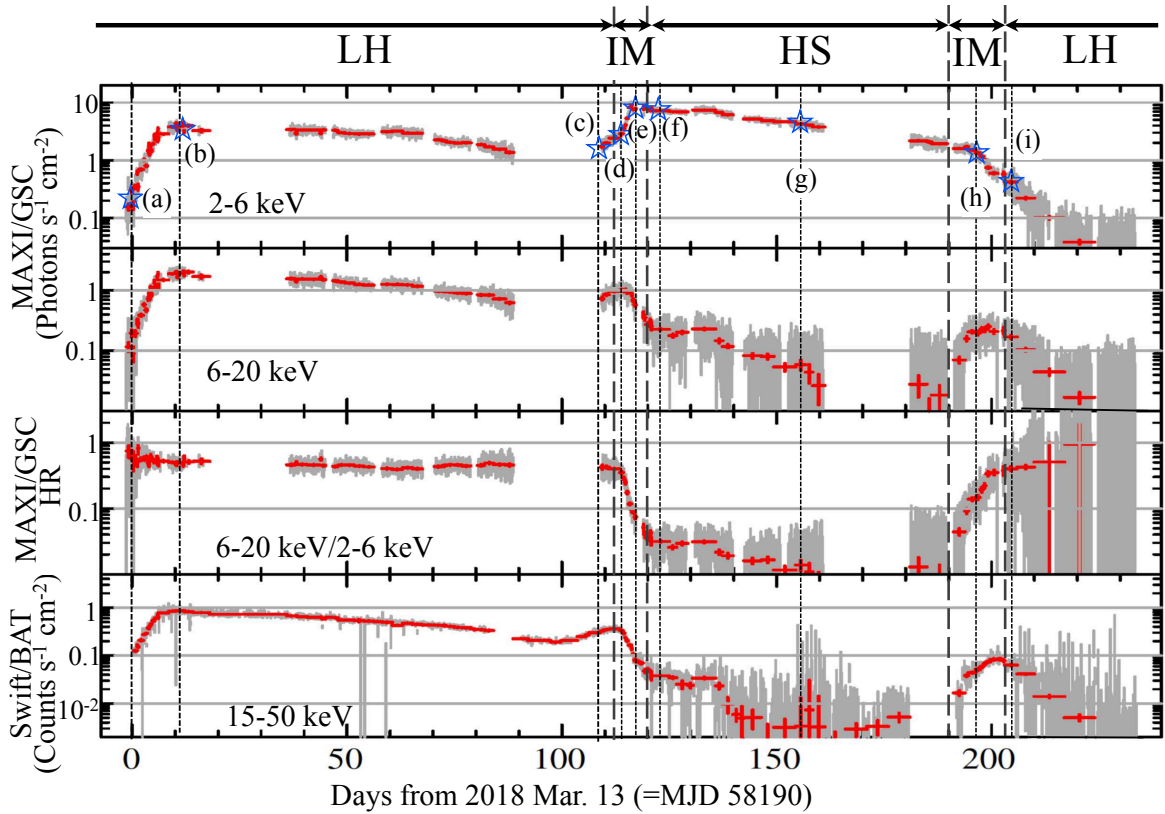}
\caption{\maxi/GSC light curves of MAXI J1820$+$070 in 2--6 keV, 
6--20 keV, their hardness ratio, and \swift/BAT light curve 
in 15--50 keV downloaded from the ``BAT Transient Monitor'' 
website \citep[][\url{http://swift.gsfc.nasa.gov/docs/swift/results/transients}]{kri13},
from top to bottom. The grey and red points show 
the data with orbital time bins ($\sim$92 min) and 
the binned data, respectively. The error bars indicate
1$\sigma$ statistical errors. The periods 
"LH", "IM", "HS" represent the low/hard, intermediate, 
and high/soft state periods, respectively. The spectra 
in the bin (a)--(i) indicated with stars 
and thin vertical lines are plotted in Fig.~\ref{fig:MAXIBATspec}. 
\label{fig:LC_longterm}}
\end{figure*}

Since its discovery, the source stayed in the low/hard state 
for more than 3 months, and then exhibited a rapid X-ray spectral 
softening in 2018 July \citep{hom18a}. This indicates that 
the source made the state transition from the so-called 
low/hard state, in which a strong hard X-ray component is 
observed in the X-ray spectrum, to the high/soft 
state, where the soft X-ray emission from the 
standard accretion disk \citep{sha73} dominates the 
X-ray flux. The source exhibited strong radio 
flares at the beginning of the transition \citep{bri18b}. 
During the transition, the radio to infrared flux decreased, 
suggesting quenching of the jet \citep{tet18,cas18}. 
By contrast, the optical flux behaved in an opposite 
way \citep{tuc18}.
In September, an X-ray spectral hardening was observed, 
indicating that the source returned to the low/hard 
state \citep{neg18,mot18,hom18b}.

In \citet{shi18} (hereafter S18), we studied the 
X-ray properties of \srcname using 
the {\maxi} and {\swift} \citep{geh04} data obtained 
during the low/hard state before the state transition in July. 
The X-ray spectrum in this period was well explained by  
Comptonization of soft X-ray photons from the standard 
disk, in the hot ($\gtrsim$ several ten keV), radiatively inefficient 
inner flow or the corona above the disk. We also analyzed a 
simultaneous multi-wavelength spectral energy distribution taken in late 
March and found that the synchrotron emission from the jet 
was likely to contribute to the optical and 
near-infrared wavelengths, whereas in the X-ray band, 
its contribution, as well as the contributions of  
synchrotron self-Comptonization and external 
Comptonization, were likely to be negligible. 

In this article, we investigate the 
long-term X-ray spectral evolution over the entire 
outburst from March to October, using the \maxi/Gas 
Slit Camera (GSC) 
and \swift/Burst Alert Telescope (BAT) data. 
We also combine the optical photometric data 
acquired with the MITSuME 50 cm telescope 
in Akeno Observatory, to study long-term optical 
flux evolution and correlation with the X-ray flux.
We adopt the solar abundance table 
given by \citet{wil00} for the spectral analysis 
described below. Errors represent the 90\% confidence 
ranges for one parameter, unless otherwise stated.

\begin{figure}[th!]
\plotone{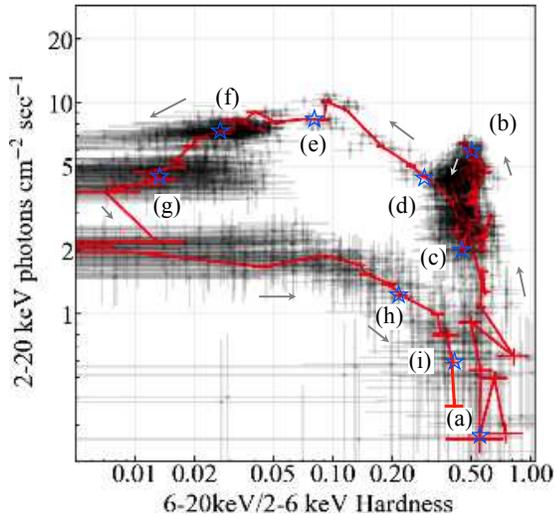}
\caption{Hardness intensity diagram of MAXI J1820$+$070, 
created from the MAXI data in Fig.~\ref{fig:LC_longterm}. 
The error bars indicate 1$\sigma$ statistical errors. \label{fig:HID}}\end{figure}

\section{X-ray Data Reduction and Analysis}
\label{sec:ana}
We reduced the $\maxi$/GSC event data 
and $\swift$/BAT survey data from 2018 March 
to October, utilizing the $\maxi$ analysis 
tools \citep{nak13} and the HEAsoft version 6.23,
and produced the light curves and spectra of 
MAXI J1820$+$070. We adopted the same 
procedures of the data reduction 
as those in S18.

\subsection{Light Curves and Hardness Ratio}
\label{sec:lc}
Figure~\ref{fig:LC_longterm} shows the \maxi/GSC 
2--6 keV and 6--20 keV light curves of MAXI J1820$+$020, 
and the {\swift}/BAT light curve in 15--50 keV 
during the 2018 outburst, 
with a time bin size of their orbital periods 
($\sim$92 minutes; grey points). The hardness 
ratio (HR) calculated from these two {\maxi} 
light curves is also plotted. 
Using the MAXI data in Fig.~\ref{fig:LC_longterm}, 
we have also plotted the hardness intensity 
diagram (HID) of MAXI J1820$+$070 in Figure~\ref{fig:HID}. 
To improve statistics, we combined the GSC data 
taken in each 1--85 adjacent orbits with similar 
fluxes into one bin (red points in Fig.~\ref{fig:LC_longterm}).

\begin{figure*}[ht!]
\epsscale{1.2}
\plotone{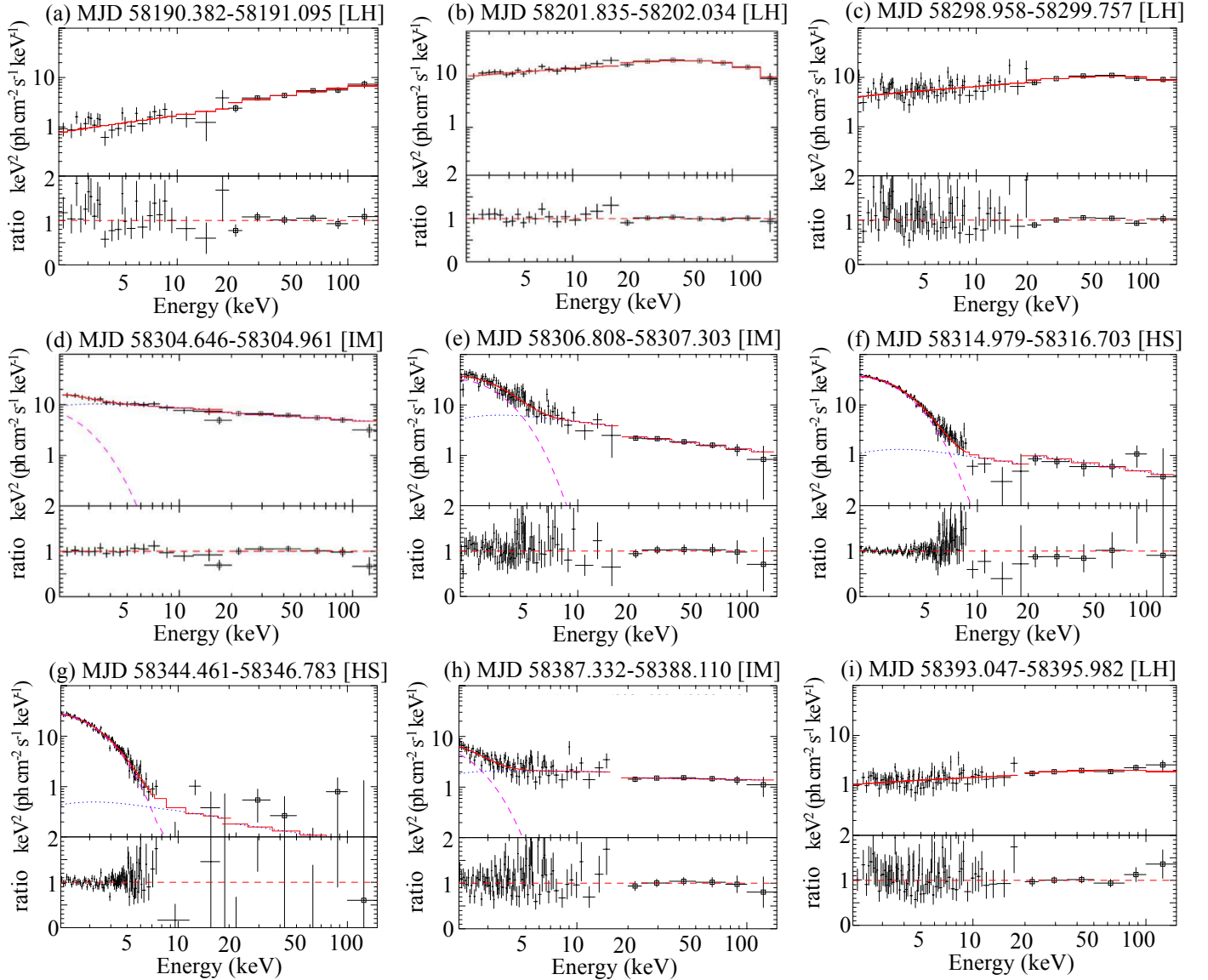}
\caption{Unfolded \maxi/GSC (cross) and \swift/BAT (open square) 
spectra, obtained at the epochs (a)--(i) indicated 
in Fig.~\ref{fig:LC_longterm} and Fig.~\ref{fig:HID}, with their best-fit models (top),  
and the data versus model ratios (bottom). 
The unscattered and scattered components are plotted in pink 
dashed and blue dotted lines, respectively, for the high/soft state and  
and intermediate state spectra.  
\label{fig:MAXIBATspec}}
\end{figure*}

The source showed a rapid flux rise and 
then a slower decay in all energy bands, 
in the first $\sim$3 months after the 
discovery on 2018 March 11. In this period, 
the HR was kept almost constant. A re-brightening 
in the 15--50 keV band was detected with the \swift/BAT 
in the middle of June (MJD $\sim$58290), whereas   
the soft X-ray data could not be obtained with 
the {\maxi} from the beginning to the end of June, 
because the source was out 
of its field of view. In early July, 
{\maxi} detected a rapid increase of the soft X-ray flux 
in 2--6 keV, to the peak level of $\sim$10 photons s$^{-1}$ 
cm$^{-2}$, corresponding to $\sim$4 Crab. Meanwhile, 
the hard X-ray flux above 6 keV decreased rapidly 
a few days before the soft X-ray peak, and the HR 
thereby dropped.

Since the second flux peak, the soft X-ray flux below 
6 keV has been decreasing gradually, whereas the 
harder X-ray flux increased again from late September 
to the beginning of October (MJD 58380--58393). 
Consequently, the HR increased again to the initial 
level before the re-brightening in June to July.
The overall evolution of the fluxes and HR 
provided a counter-clockwise track in the HID, 
like typical Galactic transient 
black hole binaries in their outburst.

Considering the behavior of the HR, 
we classified the whole outburst period into 
the canonical three spectral states of black hole X-ray binaries 
\citep[e.g.,][for details]{don07}: 
the low/hard state, the high/soft state, and the intermediate 
state (which is the transitional state between the former two states). 
\begin{description}
\item[low/hard state] {until MJD 58303.5 and from MJD 58393.0}, in which the HR was at the highest level, $\gtrsim$0.4. \item[intermediate state] {MJD 58303.5--58310.7 and MJD 58380.0--58393.0}, in which the HR was rapidly decreasing/increasing between 0.04--0.4. 
\item[high/soft state] {MJD 58310.7--58380.0}, in which the HR was lower than $\sim 0.04$.
    \end{description}
This classification is confirmed to be valid 
in the following sections, on the basis of the X-ray 
spectral profile.

\begin{deluxetable*}{cccCCCCCCC}[ht!]
\tablecaption{Best-fit 
parameters for the representative spectra \label{tab:fit}}
\tablecolumns{10}
\tablenum{2}
\tablewidth{0pt}
\tablehead{
\colhead{ID\tablenotemark{a}} & \colhead{state} &
\colhead{model\tablenotemark{b}\tablenotemark{c}} & \colhead{$\Gamma$} & \colhead{$kT_\mathrm{e}$} & \colhead{$F_\mathrm{scat}$} &
\colhead{$kT_\mathrm{in}$} & \colhead{$r_\mathrm{in}$\tablenotemark{d}} 
& \colhead{$\chi^2/\mathrm{dof}$} & \colhead{$F_X$\tablenotemark{e}} \\
\colhead{} &
\colhead{} &
\colhead{} &
\colhead{} &
\colhead{keV} &
\colhead{} &
\colhead{keV} &
\colhead{km} &
\colhead{} &
\colhead{$10^{-8}$ erg s$^{-1}$ cm$^{-2}$}
}
\startdata
(a) & 
low/hard & {\tt nthcomp} & 1.55 \pm 0.08 & >42 & -
& 0.1~\mathrm{(fixed)} & - & 29/28 & 1.6^{+0.4}_{-0.3} \\
(b) &
low/hard & {\tt nthcomp} & 1.72 \pm 0.05 & 41^{+11}_{-7} & -
& 0.1~\mathrm{(fixed)} & - & 27/27 & 14 \pm 1 \\
(c) & 
low/hard & {\tt nthcomp} & 1.74^{+0.08}_{-0.07} & 61^{+139}_{-22} & -
& 0.1~\mathrm{(fixed)} & - & 80/72 & 5.0^{+0.6}_{-0.5} \\
(d) & 
intermediate & {\tt simpl*diskbb} & 2.20 \pm 0.08  & - & 0.4^{+0.1}_{-0.2}
& 0.5 \pm 0.2 & 81^{+190}_{-40} & 18/23 & 7.5^{+1.1}_{-0.7} \\ 
(e) & 
intermediate & {\tt simpl*diskbb} &  2.4 \pm 0.2 & - & 0.10^{+0.03}_{-0.02}
& 0.67^{+0.06}_{-0.05} & 67^{+17}_{-12} & 75/80 & 10.0^{+0.7}_{-0.8} \\
 (f) &  
high/soft & {\tt simpl*diskbb} &  2.5~\mathrm{(fixed)} & - & 0.022 \pm 0.004
& 0.70 \pm 0.01 & 61 \pm 2  & 118/115 & 8.5 \pm 0.2 \\
 (g) & 
high/soft & {\tt simpl*diskbb} &  2.5~\mathrm{(fixed)} & - & 0.01 \pm 0.05
& 0.64 \pm 0.02 & 66 \pm 5  & 99/89 & 5.9 \pm 0.2 \\
 (h) & 
 intermediate & {\tt simpl*diskbb} &  2.1 \pm 0.2 & - & 0.12 \pm 0.04
& 0.45 \pm 0.08 & 73^{+61}_{-27} & 88/80 & 2.5^{+0.4}_{-0.3}\\
 (i) & 
 low/hard & {\tt nthcomp} & 1.83 \pm 0.08 & >84 & -
& 0.1~\mathrm{(fixed)} & - & 69/77 & 1.1 \pm 0.1 \\
\enddata
\tablenotetext{a}{Data IDs in Fig~\ref{fig:MAXIBATspec}.}
\tablenotetext{b}{The {\tt TBabs} model was combined for all models, with $N_\mathrm{H}$ of $1.5 \times 10^{21}$, to account for the interstellar absorption.}
\tablenotetext{c}{The seed spectrum for the Comptonization component was assumed to be a disk blackbody.}
\tablenotetext{d}{Inner radius estimated from the total photons of the disk blackbody emission, including the Comptonized photons. A distance and 
an inclination angle of 3 kpc and 60$^\circ$ are assumed, respectively. 
The color-temperature correction and the correction of the inner boundary 
condition are not considered here.}
\tablenotetext{e}{Unabsorbed 1--100 keV flux.}
\end{deluxetable*}

\subsection{Spectral Analysis}
\label{sec:spec}

We studied X-ray spectral evolution of \srcname 
using the \maxi/GSC and \swift/BAT survey data. 
We produced {\maxi} spectra in the individual 
bins presented in Fig.~\ref{fig:LC_longterm} as 
red points. Following S18, we then matched them 
with the BAT survey spectra of the individual 
continuous scans. We thus obtained 93 
simultaneous \maxi/GSC and \swift/BAT spectra 
covering from 2 keV to $\sim$200 keV over the 
whole outburst. We presented 
9 representative spectra in Figure~\ref{fig:MAXIBATspec}. 

The simultaneous \maxi/GSC and \swift/BAT spectra were 
systematically analyzed on XSPEC version 12.10.0. 
We varied the cross-normalization factor of the BAT data 
with respect to the GSC data, to take into account 
an uncertainty in the instrumental 
cross-calibration and that caused by flux variation due 
to the slight differences of the observation periods 
between the \maxi/GSC and \swift/BAT. 
We adopted a canonical spectral 
model for black hole binaries in outbursts: 
disk blackbody emission and its thermal or non-thermal 
Comptonization component. 
The {\tt TBabs} model \citep{wil00} was also employed 
to account for the interstellar absorption. We 
fixed the equivalent hydrogen column density 
$N_\mathrm{H}$ at $1.5 \times 10^{21}$ cm$^{-2}$, 
which was determined from the data of 
{\it the Neutron star 
Interior Composition Explorer} ({\it NICER}, \citealt{utt18}). 

As investigated in S18, before the spectral 
softening, the source showed a typical spectral profile 
of black bole binaries in the low/hard state: a hard 
power-law shaped spectrum with a photon index of 
$<2$, often accompanying an exponential cutoff (see 
Fig.~\ref{fig:MAXIBATspec}a--c). A similar spectrum 
was observed after the spectral hardening around 
MJD 58380--58390 (see Fig.~\ref{fig:MAXIBATspec}i). 
We applied the same model as S18, 
{\tt TBabs*nthcomp} to the spectra in 
these periods. 
The {\tt nthcomp} model calculates thermal 
Comptonization spectrum using a photon index 
and an electron temperature \citep{zyc99}.
Following S18, the seed spectrum 
was assumed to be a disk blackbody, with 
a fixed inner temperature of 0.1 keV, as 
it cannot be determined with the MAXI$+$BAT data 
covering only above 3 keV. 
The results do not depend on this value, 
as all the parameters we obtain 
remain within their 90\% error ranges, 
even when $T_\mathrm{in} =  0.5$ keV and 
0.05 keV are adopted. We ignored 
the direct disk blackbody component in $P_\mathrm{H}$.

In the high/soft state, the spectrum of \srcname was dominated by the 
soft X-rays (see Fig.\ref{fig:MAXIBATspec}f and g) 
likely originating in the thermal emission 
from the standard accretion disk \citep{sha73}. We used {\tt diskbb} 
\citep{mit84} to model this component. 
A weak power-law tail was also often observed 
above $\sim$10 keV, without a significant cutoff. 
To account for this component, we employed {\tt simpl} 
\citep{ste09}, 
assuming the non-thermal Comptonization of the disk 
emission as the origin of the hard tail. Thus, the 
model applied for the high/soft state periods 
was expressed as {\tt TBabs*simpl*diskbb}. 
The {\tt simpl} model makes a fraction of an input seed 
spectrum scatter into a power-law component, using a 
scattered fraction $F_\mathrm{scat}$ and a photon index
$\Gamma$ as the free parameters. We fixed $\Gamma$ at 2.5, 
a typical value for the high/soft state \citep[e.g.,][]{ebi94,mcc06}, 
because we were unable to constrain it, due to the low statistics 
of the \swift/BAT data. We extended the energy range 
over which the model spectrum is calculated on XSPEC, 
from those in the \maxi/GSC and \swift/BAT response 
matrix files to 0.1--500 keV, for accurate calculation 
of the convolution model {\tt simpl}.

The spectrum changed drastically during the intermediate states,  
as shown in Fig.~\ref{fig:MAXIBATspec}(c)--(e) and (h).
We applied the same model as that used for the 
spectrum in the high/soft state.  
{\tt TBabs*simpl*diskbb}, varying the photon index. 
We note that the {\tt simpl} model does 
not consider the radial dependence of the 
scattering fraction.
This means that it assumes a specific geometry 
(a disk-like geometry) for the Comptonized region, 
and its parameters may become unreliable when a 
different geometry is considered, such as a spherical 
electron cloud inside a highly truncated standard disk.
We have also tested the disk and its thermal Comptonization model, 
{\tt TBabs*(diskbb+nthcomp)}, linking the inner disk 
temperatures of {\tt diskbb} and {\tt nthcomp}, 
and have obtained equally good fit with parameter 
values similar to {\tt TBabs*simpl*diskbb}. The spectral 
cutoff was not detected significantly and only a lower limit 
of the electron temperature was obtained ($kT\mathrm{e} \gtrsim$ 50 keV).

\begin{figure*}[ht!]
\plotone{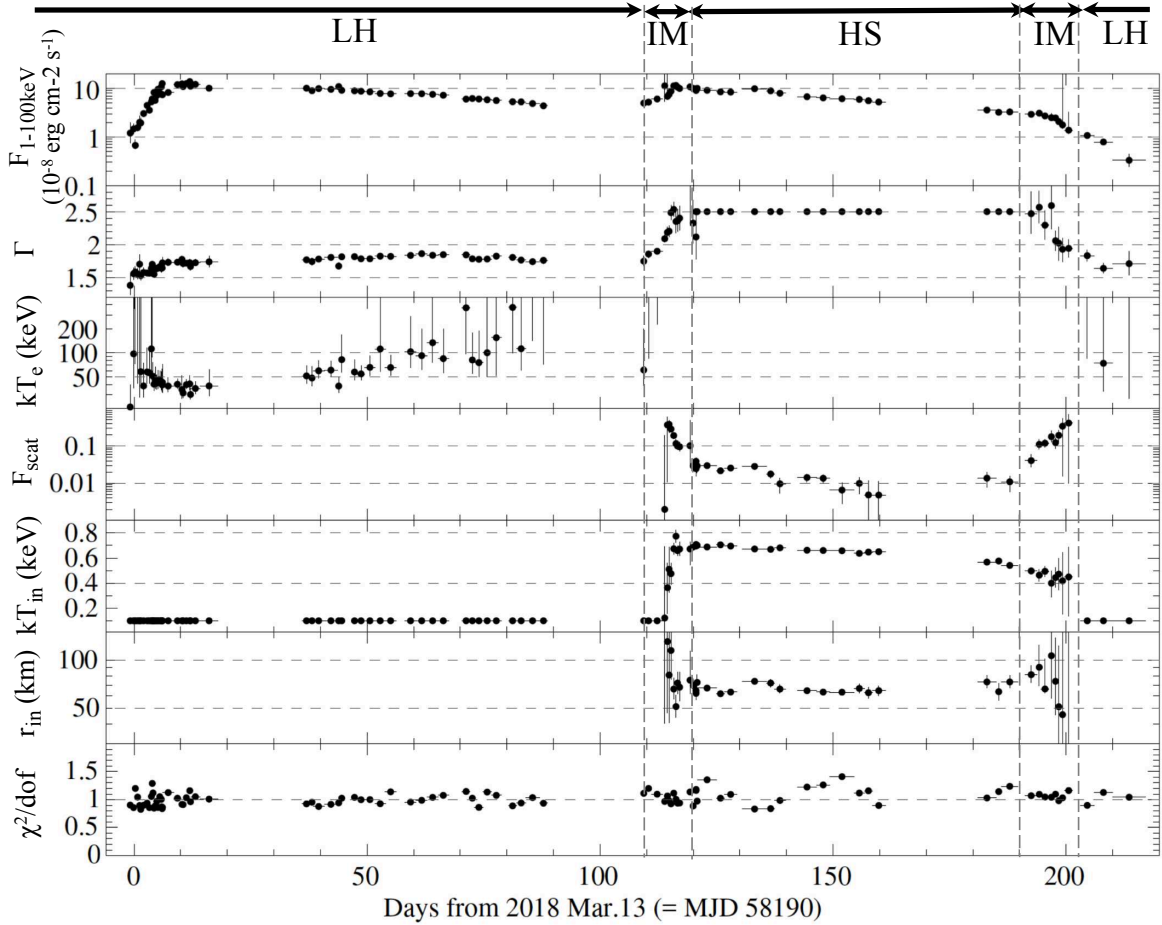}
\caption{Time variations of the parameters in the best-fit 
models. The unabsorbed 1--100 keV flux 
(in units of $10^{-8}$ erg s$^{-1}$ cm$^{-2}$, estimated 
by adopting the normalization of the \maxi/GSC data), 
the photon index, the electron temperature (keV), 
the fraction of the Comptonization component, the inner 
disk temperature (keV), the inner disk radius (km) estimated by 
assuming a distance of 3 kpc and an inclination angle of 
$60^\circ$, and the reduced chi-squared, 
from top to bottom. \label{fig:trend_fitpars}}
\end{figure*}

These models successfully described the wide-band 
{\maxi}/GSC and {\swift}/BAT X-ray spectra in the 
individual states. The best-fit models 
for the 9 representative spectra and their 
parameters are presented in Figure~\ref{fig:MAXIBATspec} 
and Table~\ref{tab:fit},respectively. 
Figure~\ref{fig:trend_fitpars} shows the trend of 
the fit parameters over the entire outburst. 
As found by S18, 
the photon index and the electron temperature slightly 
increased and decreased, respectively, during the 
flux rise at the beginning of the outburst. 
In the second flux rise from MJD$\sim$58300, 
the {\tt diskbb} component emerged from the low edge 
of the {\maxi} energy band and it contributed more to the X-ray 
spectrum as the source became brighter. In the high/soft state,  
the total X-ray flux varied by a 
factor of $\sim 5$ and the inner disk temperature varied 
between $\sim$ 0.4 and $\sim$ 0.7, whereas the inner disk 
radius was fairly constant at $\sim$60 ($D$/3 kpc) 
($\cos i/\cos 60^\circ$) km, where $D$ and $i$ are 
the distance and the inclination angle.

The iron K$\alpha$ emission line, often seen in 
other sources during the low/hard state and the intermediate 
state, was not detected significantly in any of the above spectra. 
This is most likely because of statistics and energy resolution 
of the MAXI/GSC. Fitting the spectrum (b) in Fig.~\ref{fig:MAXIBATspec}  
with the {\tt tbabs*nthcomp} model plus a Gaussian component, 
which represents the iron line, we estimated the upper 
limit on the equivalent width of the line as $\sim$300 eV. 
This is consistent with the typical values in the low/hard 
state $\lesssim 100$ eV \citep[e.g.,][]{mak08,kol14}.
In this fit, the line-center 
energy and the width were fixed to $6.4$ keV and $100$ eV, 
respectively, referring to the values obtained in 
GX 339$-$4 \citep{shi11b}. The results were confirmed to 
be unaffected, even when the line center energy is changed to 
6.9 keV, corresponding to the H-like iron, and when the line 
width is varied by a factor of 5.

\section{Optical Data Reduction and Analysis}

Optical photometric observations of MAXI J1820$+$070 
in the $g'$ band were performed with the MITSuME 50\,cm telescope 
in Akeno Observatory, Yamanashi, Japan 
\citep{kotani2005,yatsu2007,shimokawabe2008}, based on a 
Target-of-Opportunity program in the Optical and 
Infrared Synergetic Telescopes for Education and 
Research (OISTER). We reduced the $g$'-band data 
following standard procedures including 
bias and dark subtraction, flat fielding, and bad pixel masking, 
and then performed photometry of \srcname using IRAF. 
Magnitude calibration was carried out by using the magnitudes 
of nearby reference stars, which were taken from the UCAC4 catalog \citep{2013AJ....145...44Z}.
The source was also observed 
through several different optical and near-IR filters, 
with telescopes involved in the OISTER. The complete 
datasets and their analysis will be presented by 
Adachi et al. (in preparation).

\begin{figure}[ht!]
\plotone{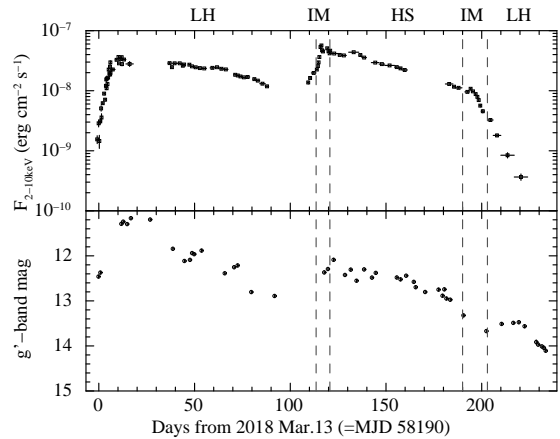}
\caption{X-ray 2--10 keV (top) and optical $g'$-band (bottom) 
light curves of MAXI J1820$+$070. The X-ray light curve was 
obtained from the best-fit model in Section~\ref{sec:spec} 
and corrected for the interstellar absorption.
The optical magnitudes were corrected for reddening 
and are expressed in the AB system. 
The error bars of the optical light curves include both 
the statistical photometric errors of MAXI J1820+070 
and the systematic errors of the reference star magnitudes 
($\sim$ 0.034 mag in total). 
\label{fig:xoptlc}}
\end{figure}

Figure~\ref{fig:xoptlc} shows the unabsorbed 2--10 keV light 
curve obtained from the best-fit models in Section~\ref{sec:spec}, 
and the optical $g'$-band light curve from the MITSuME telescope. 
The optical data were dereddened 
with the reddening factor $A_{g'} = 0.98$. This value was obtained 
by using the extinction law given in \citet{car89} and the 
reddening factor in the $V$-band, $A_V = 0.84$, which was 
derived from the absorption column $N_\mathrm{H} = 1.5 \times 10^{21}$ 
cm$^{-2}$ via the $N_\mathrm{H}$ vs. $A_V$ relation in \citet{pre95}.
The errors of the magnitudes include both the statistical 
photometric errors of MAXI J1820$+$070 and the systematic errors 
of the reference star magnitudes (typically $\sim$0.034 mag in total).
Overall, the optical flux increased (decreased) with increasing 
(decreasing) X-ray flux, although a weak optical re-brightening took 
place just after the soft-to-hard transition during the X-ray decay (MJD $\sim$58400), 
as was reported by \citet{bag18b}.
A remarkable point is that the $g'$-band flux at the first peak 
on MJD $\sim$58202 is higher than that at the second peak on 
MJD $\sim$58306 by $\sim 1$ mag (by a factor of $\sim 2.5$ in 
the flux unit), whereas the X-ray flux at the first peak is  
slightly lower than that at the second peak.

\begin{figure}[ht!]
\plotone{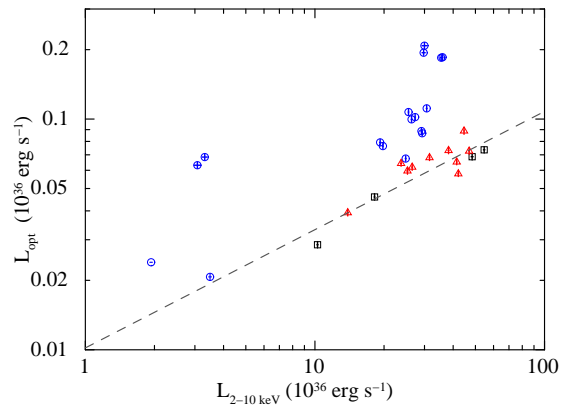}
\caption{Correlation of X-ray and optical luminosities of 
MAXI J1820$+$070. Blue circles, black squares, and red triangles 
indicate the data taken in the low/hard state, the intermediate state, 
and the high/soft state, respectively. The dashed lines are 
the best-fit power-law function for the data points 
in the latter two states,  
$L_\mathrm{opt} \propto 
L_\mathrm{2-10~keV}^{0.51 \pm 0.03}$.
\label{fig:xoptcor}}
\end{figure}

To investigate a flux correlation between 
the X-ray and optical bands, we also plotted the 
optical vs. X-ray luminosity diagram in 
Figure~\ref{fig:xoptcor} using the simultaneous 
data in Fig.~\ref{fig:xoptlc}. 
The optical magnitudes were converted to fluxes, 
and multiplied by the central frequency of the 
$g'$-band filter, $\nu = 6.25 \times 10^{14}$ 
Hz, following \citet{rus06}, to obtain the fluxes in the $\nu F_\nu$ units. Using these extinction/absorption-corrected 
optical and X-ray fluxes, we estimated the optical and 
X-ray luminosities, $L_\mathrm{opt}$ and $L_\mathrm{2-10~keV}$, 
respectively, assuming an isotropic radiation at a distance of 3 kpc.  
As is noticed in this figure, the data points are more highly 
scattered and $L_\mathrm{opt}$ with respect to $L_\mathrm{2-10~keV}$ 
is somewhat higher in the low/hard state than in 
the intermediate state and the high/soft states. The 
data points in the latter two states can be 
described with a power-law function. 
Fitting the data of the two states in Fig.~\ref{fig:xoptcor} 
together, we obtained the best-fit model of $L_\mathrm{opt} \propto 
L_\mathrm{2-10~keV}^{0.51 \pm 0.03}$.

\section{Discussion}
\subsection{Interpretation of the Long-term X-ray Evolution} \label{sec:discussion_lcspec}

We have monitored \srcname in 2--200 keV during its outburst
using \maxi/GSC and \swift/BAT. In the soft X-ray band below 
6 keV, it showed two ``rapid rise and slow decay''-type 
long-term flux evolutions (hereafter we call them 
as the first/second sub-outbursts). This profile 
is often seen in X-ray light curves of 
transient black hole X-ray binaries.
The two big sub-outbursts suggest that, in this outburst, 
the source 
exhibited two discrete increases in 
its mass accretion rate through the disk. Re-flares in the outburst 
decay have been observed in many other black hole X-ray 
binaries, and they usually occur after 50--100 days 
from the first flux peak \citep[e.g.,][]{che93, nak14}. 
This interval is similar to that of the two sub-outbursts 
in MAXI J1820$+$070.

Such re-flares were often 
interpreted by the enhancement of the mass 
accretion rate from the donor star, which 
was inflated due to the X-ray irradiation 
\citep{che93,min94}. 
This could also explain the second sub-outburst 
of MAXI J1820$+$070, as was discussed in S18.
However, the re-flares in other sources are 
usually much weaker than their main 
outbursts, having $\sim$1 order of magnitude 
lower peak luminosities than those of 
the main outbursts. Why such a strong 
re-brightening took place in MAXI J1820$+$070
is an open question, but maybe it is because 
the source stayed in the low/hard state during 
the first sub-outbursts, and the strong hard 
X-rays heated the companion star more efficiently 
than other sources with re-flares, which often 
show state transitions in their main outbursts.

The HR was almost constant at a high level 
during the first sub-outburst, whereas 
its rapid decrease was observed in 
the second sub-outburst. This indicates that the state 
transition from the low/hard state to the 
high/soft state (the hard-to-soft transition) 
took place only in the second 
sub-outburst. Interestingly, the 1--100 keV fluxes 
were almost the same (different only by $\sim$10\%) 
at the flux peaks in the two sub-outbursts. 
If the radiation efficiency was constant 
during the outburst, 
this suggests that the mass accretion rates were 
the same in the two peaks, and that the hard-to-soft 
transition is not determined the mass accretion rate 
alone. We note that, however, in reality, the radiation 
efficiency most likely varies, depending 
on the properties of the accretion flow and outflows 
\citep[e.g.,][and references therein]{mal04,kat08}.
Hence, the same X-ray luminosity does not necessarily 
mean the same mass accretion rate, and it is difficult 
to completely rule out, using our results alone, 
the possibility that the mass accretion rate at 
the beginning of the hard-to-soft transition 
was higher than that at the first X-ray peak.

More direct evidence that something 
other than the mass accretion rate plays a
role to trigger the hard-to-soft transitions, 
has been found in previous observations. 
As was observed in other sources such as GX 339$-$4, 
the hard-to-soft transitions actually took place 
at different luminosities in different outbursts 
of the same sources \citep[e.g.,][]{dun08}.
What made the difference in the two sub-outbursts 
of \srcname is still unclear, but it may be related 
to the magnetic fields in the inner disk region, 
considering previous theoretical studies of 
the accretion flows in the hard-to-soft transition 
\citep{mat06, beg14}. Perhaps, in the first sub-outburst, 
strong magnetic fields were present in the inner disk 
region for some unknown reasons, 
and prevented from developing the standard disk inwards. 

Although the low/hard state period before the hard-to-soft 
transition was unusually long, 
the combination of two sub-outburst provided a normal ``q''-shaped 
track in the HID, like other black hole binaries. 
The hysteretic behavior in the HID, where the hard-to-soft 
transition occurs at higher luminosity than that of the 
transition from the high/soft state back to the low/hard 
state (the soft-to-hard transition), 
is seen not only in black hole binaries but 
also neutron star low mass X-ray binaries \citep{mun14}. 
However, MAXI J1820$+$070 is unlikely to harbor 
a neutron star. 
Typical soft X-ray spectra of neutron star 
low-mass X-ray binaries at high luminosities 
show a thermal emission component with a temperature 
of $\sim 1$ keV and a comparably strong Comptonization 
component located at slightly higher energies than 
the thermal component, and these two components are 
considered to originate in the blackbody emission of the 
neutron star surface and a Comptonized disk emission 
\citep{mit89}, respectively, or the disk blackbody emission 
and a Comptonized neutron star blackbody emission \citep{chu95}, 
respectively. By contrast, all the high/soft-state spectra 
of MAXI J1820$+$070 below 10 keV were successfully described by a 
single disk blackbody model with $T_\mathrm{in} \lesssim 1$ keV.
We have confirmed that the fit of the spectrum (f) in 
Fig.~\ref{fig:MAXIBATspec} is not improved when 
{\tt nthcomp} with $kT_\mathrm{e} = 5$ keV (a typical value 
for the Comptonization component of neutron star X-ray binaries) is 
added to the {\tt simpl*diskbb} model. These results support 
that its compact object is not a neutron star but a black hole.

The all-time monitoring and the wide coverage 
provided by the combination of the \maxi/GSC and 
\swift/BAT, have enabled us to simultaneously 
determine the levels of the disk and Comptonization 
components in the X-ray spectrum of MAXI J1820$+$070, 
and study their long-term variations in the outburst. 
We discuss the evolution of the structure 
of the accretion flow, based on the results 
from our spectral analysis.

In the low/hard state, the spectrum was well described 
with a thermal Comptonization 
component with $\Gamma < 2.0$ and $kT_\mathrm{e} \gtrsim$ 
40 keV. We detected a gradual increase of 
$\Gamma$ and decrease of $kT_\mathrm{e}$ 
during the rapid flux increase in the first sub-outburst 
(see Tab.~\ref{tab:fit} and Fig.~\ref{fig:trend_fitpars}).
As discussed in S18, this could be explained 
by the disk truncation model, in which the 
standard disk is truncated in the 
beginning of the outburst, and it develops 
inward as the X-ray flux increases
\citep[e.g.,][for more details]{don07}. 
The seed photons for Compton scattering 
increases as the standard disk extends 
inward, which would cause a more effective 
cooling of the hot inner flow or corona. 

In the intermediate state, where the spectral 
softening/hardening took place, the inner disk temperature 
rapidly increased as the total X-ray flux increased, 
while the fraction of the Comptonization component 
in the total X-ray flux dropped  
and the photon index increased up to $\sim$ 2.5. 
The rapid change of the X-ray spectrum 
is typical of black hole X-ray binaries 
in the intermediate state. In this phase, $r_\mathrm{in}$ 
was not well constrained, due to uncertainties 
in the inner disk temperature and the normalization 
of the direct disk component, whose contribution 
to the total X-ray flux was smaller than that 
in the high/soft state.

After the hard-to-soft transition, the contribution of 
the Comptonization component decreased down to less than 
a few \%, a typical value in the high/soft state.
During the high/soft state, 
the inner radius was almost constant 
at $r_\mathrm{in} \sim 60$ ($D$/3 kpc) ($\cos i/\cos 60^\circ$)$^{-1/2}$ km, 
even though the 1--100 keV flux decreased 
by a factor of $\sim$ 5. This indicate that the 
standard disk extended stably down to the innermost 
stable circular orbit (ISCO) in this period.

\begin{figure}[th!]
\plotone{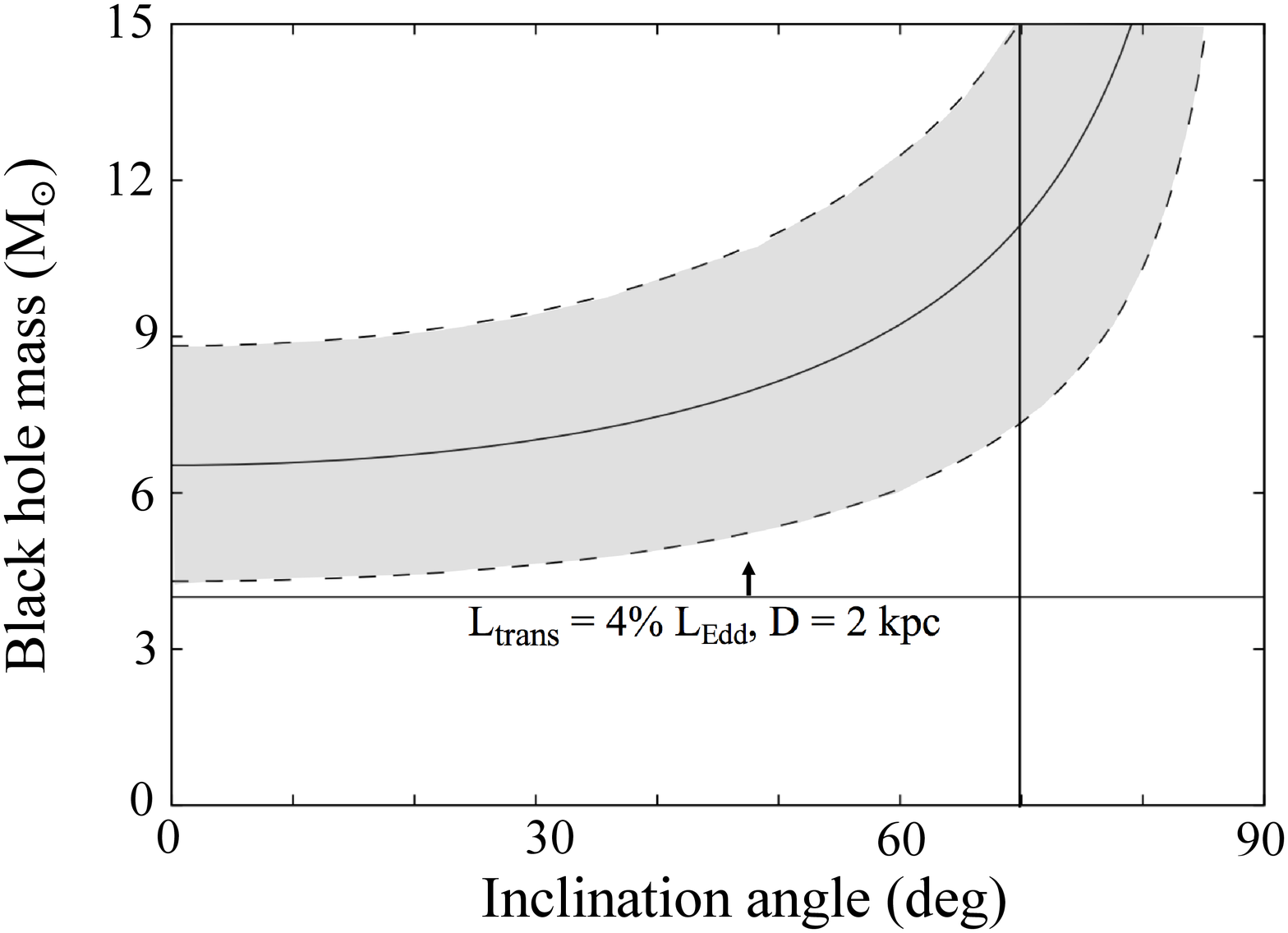}
\caption{Constraints on the black hole mass 
and the inclination angle of MAXI J1820$+$020, 
in the case of a non-spinning black hole. The 
horizontal and vertical lines 
present the lower limit of the mass imposed 
by the luminosity at the soft-to-hard 
transition, and the lower limit of the 
inclination angle at which the dips 
are observed, respectively. 
The solid curve shows the relations 
obtained from the constant inner radius 
in the high/soft state, and the 
dashed ones separating the 
shadowed region are its upper/lower 
limits considering the errors of 
the inner disk radius and the distance.
\label{fig:M_i}}\end{figure}

\subsection{Estimation of System Parameters} \label{sec:discussion_mass}

We have attempted to constrain the black hole mass 
using the results of our spectral analysis. 
\citet{mac03} suggested that the soft-to-hard 
transition occurs at 1-4 \% Eddington luminosity 
($L_\mathrm{Edd}$). The 0.01--100 keV flux  
at the beginning of the soft-to-hard transition is 
estimated to be $\sim 5 \times 10^{-8}$ erg cm$^{-2}$ 
s$^{-1}$ from the best-fit model, which 
can be converted to a luminosity of $\sim 5 \times
10^{37}$ ($D$/3 kpc)$^2$ erg s$^{-1}$, if an isotropic 
emission is assumed. 
Considering the uncertainties in the transition 
luminosity and the distance ($\approx 3 \pm 1$ kpc; \citealt{gan18b}), 
the lower limit of the black hole mass ($M_\mathrm{BH}$)
of \srcname is calculated as 4 $M_\sun$, when 
$D=2$ kpc and the transition 
luminosity of $4$\% $L_\mathrm{Edd}$ are adopted.
This value justifies that the source has a black hole accretor.

Another constraint is given 
from the inner disk radius in the high/soft state, 
which is considered to be identical the radius of 
the ISCO. We derive $r_\mathrm{in} = 66.0 \pm 0.8$ 
($D$/3 kpc) ($\cos i/\cos 60^\circ$)$^{-\frac{1}{2}}$
km as a weighted average of the inner radii estimated 
in the high/soft state. 
Applying the combined correction factor of 1.18, 
for the stress-free boundary 
condition and the color-temperature correction 
\citep[see e.g.,][]{kub98}, 
we obtain $R_\mathrm{in} = 77.9 \pm 1.0$ 
($D$/3 kpc) ($\cos i/\cos 60^\circ$)$^{-\frac{1}{2}}$ km. 
In Figure~\ref{fig:M_i}, we plot the constraint 
obtained by equating this radius to that of the ISCO,  
as a shaded region on the $M_\mathrm{BH}$ 
versus $i$ plane. Here, we have assumed a non-rotating 
black hole, in which case th radius of the ISCO 
is 6 $GM_\mathrm{BH}/c^2$ 
(where $G$ and $c$ are the gravitational constant and 
the light speed, respectively).

In the beginning of the outburst, {\it NICER} detected a 
strong absorption dips \citep{hom18c}, which are often seen 
in X-ray binaries with high inclination angles 
above $\sim 70^\circ$ \citep[e.g.,][and references therein]{boi03, hom05}. 
If \srcname has a high inclination angle similar to other dipping 
sources, a lower limit of $M_\mathrm{BH}$, $\sim 
7 M_\sun$, is obtained from Fig.~\ref{fig:M_i}. 
The origin of the dips in \srcname is not yet fully 
understood, however; the dips have not been detected 
after the flux peak in the first sub-outburst,
Also, ionized absorption lines from disk winds, 
also often observed in high inclination sources, 
have never been detected in the outburst. The allowed 
range of $M_\mathrm{BH}$ becomes lower if the 
source has a smaller inclination angle than other 
X-ray binaries with dips and ionized absorption lines. 

If the black hole of \srcname is a Kerr (rotating) 
black hole, the $M_\mathrm{BH}$ values 
become larger (by up to a few to several 
times in the case of the maximum spin) than 
those in the zero-spin case, 
primarily because the ISCO radius decreases 
from 6 $GM_\mathrm{BH}/c^2$ up to $\sim 1 GM_\mathrm{BH}/c^2$,
as the black hole spin becomes higher 
(see \citealt{dav11} for more detailed discussion and 
\citealt{dav11, shi11a, wan18} for comparisons of 
the constraints on $M_\mathrm{BH}$ among different spin parameters).
In the case of a highly spinning black hole, 
the strong relativistic effects, 
including gravitational redshift and light 
bending in the vicinity of the black hole, 
and the Doppler beaming due to the disk rotation, 
modifies the observed disk spectrum from that obtained 
with the simple {\tt diskbb} model, and hence 
disk emission models considering these effects, 
instead of the simple multi-color disk model, 
must be employed to describe the spectra and to determine 
$M_\mathrm{BH}$ accurately. 
However, the {\maxi} spectra do not have statistics 
and soft X-ray coverage sufficient to apply such 
relativistic disk emission models. For further studies, 
we need data from soft X-ray detectors like $NICER$, with high 
sensitivity but free from the pile-up effects, 
This is beyond the scope of this work and we 
leave it as a future work. 

Also, the common transition luminosity estimated 
by \citet{mac03} may not always be applicable; 
an exception has recently been found in the 
outburst of MAXI J1535$-$571 \citep{nak18}, 
which stayed in the high/soft state even after 
the X-ray flux decreased $\sim$ 3 orders of 
magnitude from the luminosity of the hard-to-soft 
transition. To derive more accurate $M_\mathrm{BH}$
and to test the argument on the transition 
luminosity, determination of the mass function 
of \srcname is required via optical spectroscopy 
in the quiescence phase. 

\subsection{Optical and X-ray Correlation}

In most of the outburst period, the optical flux 
increased/decreased with the X-ray flux 
rise/decay, although in the optical band, 
the second sub-outburst
was weaker than the first 
sub-outburst, unlike those in the X-ray 
band; in 2--10 keV, the second peak flux 
was twice higher than that of the first peak. 
Consequently, the $L_\mathrm{opt}/L_\mathrm{2-10~keV}$ 
ratio in the intermediate state and the high/soft state  
is relatively small compared with that in 
$P_\mathrm{H}$ (see Fig.~\ref{fig:xoptcor}). 
Similar trend was found in other BHBs 
\citep{rus06}. The difference in the 
$L_\mathrm{opt}/L_\mathrm{2-10~keV}$ 
value can be explained by the contribution 
of the jet in the optical band during the low/hard 
state. Indeed, S18 found that the jet emission 
can substantially contribute to the optical 
flux but not significantly to the X-ray flux of 
MAXI J1820$+$070 during the low/hard state. 
A smaller $L_\mathrm{opt}/L_\mathrm{2-10~keV}$ 
value would be obtained if the jet in  
suppressed in the other states, as suggested  
by \citet{tet18,cas18}. 
We note, however, that the synchrotron emission 
from a magnetized hot inner accretion flow 
could also considerably contribute to the optical 
flux and its variation in the low/hard state, 
as suggested by \citet{vel18}. 
It is difficult to separate the contributions of 
the jets and the hot inflow, because they 
can have similar SED profiles and variability 
properties \citep{vel13, vel18}.

We have found a tight correlation between 
the X-ray and optical luminosities in the 
intermediate state and the high/soft state,
which is well reproduced by a power-law model 
as $L_\mathrm{opt} \propto 
L_\mathrm{2-10~keV}^{0.51 \pm 0.03}$, 
whereas the data points in the low/hard state 
are strongly scattered. Considering that 
our optical and X-ray observations were 
not exactly simultaneous, the scatter 
in the low/hard state 
could be produced 
by the strong flux variations in both 
optical and X-ray bands on timescales of 
$<$day, which are likely to originate in  
synchrotron radiation from the jets 
or the hot inflow, and Comptonization of 
the disk photons in
the hot inflow/corona, respectively 
(e.g., \citealt{gan18a}; \citealt{tuc18}; S18). 
By contrast, appearance of the strong X-ray-optical 
correlation in the intermediate state and the high/soft state 
would indicate the suppression of such variability 
in these states, and this could be explained if 
the jets were quenched and the standard disk 
was developed close to the ISCO, replacing the 
hot inner flow.

When the variable emission components are suppressed,  
the blackbody emission from the companion star, and/or 
the reprocessed emission from the outer disk 
irradiated by strong X-rays,
are the possible origins of the the optical flux.
The former possibility is unlikely in MAXI J1820$+$070, 
because its optical magnitude in quiescence was 
found to be $\gtrsim 18$ mag, 
much lower than those during the outburst (S18). 
Hence, the observed correlation 
in the intermediate state and the high/soft state 
would be produced by variation of the disk emission.
If the optical band is located in the Rayleigh-Jeans (RJ) 
part of the disk spectrum, 
$L_\mathrm{opt}$ should be proportional to 
$L_\mathrm{X}^{1/4}$. If it is located 
in the flat part of the irradiated disk 
spectrum, which is produced at higher 
frequencies than the RJ part by the 
reprocessing of the X-rays \citep{gie09}, 
$L_\mathrm{opt} \propto L_\mathrm{X}$ 
is realized \citep[e.g.,][]{cor09,shi11b}.
The index of the correlation that we obtained, 
$0.51 \pm 0.03$, is larger than that expected in the RJ part 
and smaller than that of the flatter part, and therefore 
the $g'$-band frequency of MAXI J1820$+$070 
may correspond to the transition region 
between the RJ and flat parts, somewhere around 
the peak frequency of the blackbody emission 
of the outer disk edge.

If the above discussion is correct, the 
$g'$ band approximately corresponds to the 
peak of the blackbody emission from the irradiated 
outer disk edge, and this  
gives the outer disk temperature, $T_\mathrm{out,eff} \sim 0.1$ eV.
The total flux generated at the outer edge, 
$\sigma T_\mathrm{out,eff}^4$ (where $\sigma$ is 
the Stefan-Boltzmann constant) is determined by the sum of 
the intrinsic disk flux at the outer edge, 
$\sigma T_\mathrm{in}^4 \left(\frac{R_\mathrm{out}}{R_\mathrm{in}} \right)^{-3}$ 
(where $R_\mathrm{out}$ is the outer disk radius), 
and the illuminating X-ray flux that is thermalized 
at $R_\mathrm{out}$. Assuming the irradiated disk model 
constructed by \citet{gie09}, where the illuminating
X-ray flux decreases in proportion to $r^{-2}$ 
and a fraction $f_\mathrm{out}$ of it is thermalized, 
the total flux at $R_\mathrm{out}$ is expressed as 
\begin{equation}
    \sigma T_\mathrm{out,eff}^4 = \sigma T_\mathrm{in}^4 \left(\frac{R_\mathrm{out}}{R_\mathrm{in}} \right)^{-3} + f_\mathrm{out} 
    \sigma T_\mathrm{in}^4 \left(\frac{R_\mathrm{out}}{R_\mathrm{in}} \right)^{-2}.
\end{equation}
In the high/soft state, the second term is usually 
dominant and $f_\mathrm{out}$ is typically $\sim 10^{-3}$ \citep[e.g.,][]{gie09}.
Adopting $T_\mathrm{in} \sim 0.7$ keV, a typical value 
of MAXI J1820$+$070 in the high/soft state, we obtain 
$R_\mathrm{out} \sim 10^6 R_\mathrm{in} \sim 10^7$ km. 
Application of this value to the Keplar's third law 
gives a lower limit on the orbital period, 
$\sim 2~((M_\mathrm{BH}+M_\mathrm{c})/10 M_\sun)^{-1/2}$ day, 
where $M_\mathrm{c}$ is the mass of the companion star. 
The above calculation is just a rough order-of-magnitude 
estimation based on some assumptions, and should be tested in 
the future by e.g., modeling spectral energy distribution 
from optical to X-ray bands, and measuring orbital modulations 
of emission/absorption lines through optical spectroscopy.

\section{Summary}
We have studied the nature of \srcname 
and the evolution of its accretion 
disk structure, using the {\maxi}/GSC 
and {\swift}/BAT data obtained in 
almost the entire period of the 2018 outburst.
The behavior in the hardness intensity diagram, 
the basic properties of the X-ray spectra, and 
the constraints on accretor's mass obtained 
from our spectral analysis, indicate that the 
source is a black hole binary. 
We found that the outburst is composed of two 
``fast rise and slow decay''-type flux evolutions, 
with almost the same total X-ray luminosity at the 
peaks, although the state transition only took place 
in the second flux rise and decay. This implies 
that the mass accretion rate is not the unique 
factor to trigger the transition to the high/soft 
state. The X-ray spectrum was well described with 
the standard disk emission and its Comptonization, 
and its long-term variation can be explained 
in terms of the disk truncation model. The 
optical-X-ray luminosity correlation suggests 
that in the low/hard state the optical flux 
was substantially contributed by emission 
from the jets, which was likely to be suppressed 
in the intermediate and high/soft state.

\acknowledgments

We acknowledge the use of {\maxi} data provided by RIKEN, JAXA 
and the {\maxi} team, and of public data from the {\swift} 
data archive. Part of this work was financially supported 
by Grants-in-Aid for Scientific Research 16K17672 (MS), 
17H06362 (NK, YU, HN), and 16K05301 (HN) 
from the Ministry of Education, Culture, Sports, 
Science and Technology (MEXT) of Japan. 
This work was supported by Optical and Near-Infrared Astronomy 
Inter-University Cooperation Program and the joint research program 
of the Institute for Cosmic Ray Research (ICRR).

%

\vspace{5mm}
\facilities{\maxi(GSC), \swift(BAT)}

\end{document}